\documentclass[12pt]{article}
\usepackage{cite}
\usepackage{amsmath}
\usepackage{amsfonts}
\usepackage{amsthm}

\newtheorem{theorem}{Theorem}[section]

\newtheorem{lemma}[theorem]{Lemma}

\newtheorem{conjecture}[theorem]{Conjecture}
\theoremstyle{definition}
\newtheorem{definition}[theorem]{Definition}
\newtheorem{remark}[theorem]{Remark}
\newtheorem{example}[theorem]{Example}
\begin{document}
\baselineskip 0.60cm
\begin{center}
{\Large {\bf The structure of product bases of $\mathbb{C}^{2}\bigotimes\mathbb{C}^{n}$ }}
\vskip 10mm
{\bf Xilin Tang\footnote{Corresponding author. Supported in part by the NNSF of China (No. 11571119).}, Yanna Liu and Ze Gu}

{\small{\sl Department of Mathematics, South China University of Technology,\\ Guangzhou, Guangdong, 510640, P.R. China \\
{\sl E-mail: xilintang2016@sina.com, mzyanna@163.com, guze528@sina.com}}}
\vskip 10mm

\begin{minipage}{120 mm}
{\small {\bf Abstract}~~In this paper, we mainly characterize the structure of product bases of the complex vector space
 $\mathbb{C}^{2}\bigotimes\mathbb{C}^{n}$. It gives an  answer to the conjecture in
case of $d=2n$ proposed by McNulty et al in 2016. As the application of the result, we obtain all the product bases of a bipartite
system $\mathbb{C}^{2}\bigotimes\mathbb{C}^{n}$. It is helpful to review the
structure of all the product bases of $\mathbb{C}^{2}\bigotimes\mathbb{C}^{2}$ and $\mathbb{C}^{2}\bigotimes\mathbb{C}^{3}$, which given by McNulty et al.}
\end{minipage}
\end{center}
\vskip 1cm

\section{Introduction and preliminaries}

Einstein, Podolsky and Rosen (EPR) \cite{1} first highlighted the important feature of quantum mechanics which we now call \textbf{entanglement}, that is, a quantum state $|\varphi\rangle\in\mathbb{H}_{A}\bigotimes\mathbb{H}_{B}$
can not be represented in the form $|\varphi\rangle=|\varphi_{A}\rangle\otimes|\varphi_{B}\rangle$, where
 $|\varphi_{A}\rangle\in\mathbb{H}_{A}$, $|\varphi_{B}\rangle\in\mathbb{H}_{B}$. Otherwise, a quantum state
is unentangled in a bipartite system, we call it a \textbf{separable state} (or \textbf{product state}).

\begin{definition}(see \cite{2})
~A vector $|v\rangle$ in the tensor product $\bigotimes^m_{i=1}{\mathbb{C}^{{d}_{i}}}$ is called a pure product vector if it is a
 vector of the form $|{v}_{1}\rangle\otimes|{v}_{2}\rangle\otimes\cdots\otimes|{v}_{m}\rangle$, where $|v_{i}\rangle\in{\mathbb{C}^{{d}_{i}}}$, $i=1, 2, \ldots, m$.
\end{definition}

\begin{definition}{\textbf{(Definition 1 \cite{3})}}
~An orthonormal basis $B$ of a complex vector space $\mathbb{C}^{d}=\bigotimes^m_{i=1}{\mathbb{C}^{{d}_{i}}}$
with dimension $d=d_{1}d_{2}\cdots d_{m}$
is a product basis if each element in $B$ takes a pure product vector.
\end{definition}

Two orthonormal bases
$\{|a_{i}\rangle~|~i=1, 2, \ldots, d\}$ and
$\{|b_{j}\rangle~|~j=1, 2, \ldots, d\}$ of a complex vector space
$\mathbb{C}^{d}$ are \textbf{mutually unbiased (MU)}
 if $|\langle a_{i}|b_{j}\rangle|^2=\frac{1}{d}$ for
all $i, j\in\{1, 2, \ldots, d\}$. The study for MU bases is attractive in recent
years since MU bases play important roles
in quantum communications.
It is  known that the number
of MU bases of the complex vector space $\mathbb{C}^{d}$ is less than or equal to $d+1$ \cite{4}. In particular,
the maximum number can be reached if $d=p^{n}$\cite{5}, where $p$
is a prime number, $n\in\mathbb{N}_{+}$ and $\mathbb{N}_{+}$ denotes the set of positive integers.
 However,
whether the bound can be reached for a composite number $d$ is still an open problem if $d\geq 6$.
 In particular, it is not known if
 there exist more than three MUBs in dimension 6. Besides, the three MUBs in dimension 6 are
 the forms
of product bases. Based on this situation, many researchers
began to study the MU product bases (MUPBs)\cite{3,6,7}. DiVincenzo and Terhal\cite{8}
 introduced
the product bases. McNulty and Weigert\cite{6} discussed
 all the product bases
in $d=4, 6$. Clearly, a basis of subsystem $\mathbb{H}_{A}$ tensors a basis of
subsystem $\mathbb{H}_{B}$
constitutes a basis of the bipartite system $\mathbb{H}_{A}\bigotimes\mathbb{H}_{B}$.
 Naturally, we want to know
 whether the product bases of a bipartite system can be grouped into the bases of two subsystems.
 McNulty et al \cite{3} proposed a conjecture about the structure of product bases of a bipartite system as follows.

\begin{conjecture}
~The set $B=\{|a_{i},b_{i}\rangle~|~i=1, 2, \ldots, d=d_{1}d_{2}\}$ is an orthonormal product basis of
the space $\mathbb{C}^{d_{1}}\bigotimes\mathbb{C}^{d_{2}}$ if and only if the $d$ vectors $|a_{i}\rangle\in\mathbb{C}^{d_{1}}~(i=1, 2, \ldots, d)$ and the $d$ vectors
 $|b_{i}\rangle\in\mathbb{C}^{d_{2}}~(i=1, 2, \ldots, d)$ can be grouped into $d_{2}$ orthonormal bases
 of $\mathbb{C}^{d_{1}}$ and $d_{1}$ orthonormal bases of $\mathbb{C}^{d_{2}}$.
\end{conjecture}

The sufficiency is not true by the following:

\begin{example}
~Let $B=\{|{a}_{1}\rangle\otimes|{b}_{1}\rangle,
|{a}^{\bot}_{1}\rangle\otimes|{b}^{\bot}_{1}\rangle,
|{a}_{2}\rangle\otimes|{b}_{2}\rangle, |{a}^{\bot}_{2}\rangle\otimes|{b}^{\bot}_{2}\rangle\}$
 be a subset of $\mathbb{C}^2\otimes\mathbb{C}^2$,
where $rank\{|{x}\rangle,|{y}\rangle\}=2$ for all distinct $|{x}\rangle, |{y}\rangle\in
\{|{a}_{1}\rangle,|{a}_{2}\rangle,|{a}^{\bot}_{1}\rangle,|{a}^{\bot}_{2}\rangle\}$ or
$|{x}\rangle, |{y}\rangle\in
\{|{b}_{1}\rangle,|{b}_{2}\rangle,|{b}^{\bot}_{1}\rangle,|{b}^{\bot}_{2}\rangle\}$.
 Obviously, the set $B$ satisfies the condition of Conjecture 1.3.
 However, $B$ is not an orthonormal product basis of $\mathbb{C}^2\bigotimes\mathbb{C}^2$.
\end{example}

The remainder of this paper is organized  as follows. In Section 2, we characterize the structure
of product bases of $\mathbb{C}^{2}\bigotimes\mathbb{C}^{n}$ and obtain the Main theorem. This shows that  the
first components and the second components of product bases of
$\mathbb{C}^{2}\bigotimes\mathbb{C}^{n}$ can be grouped into $n$ orthonormal bases
of $\mathbb{C}^2$ and 2 orthonormal bases of $\mathbb{C}^n$ respectively, which
 answers  the modified conjecture in case of $d=2n$. In Section 3,
we review all the product bases in $d=4$ and $d=6$ as application of the Main theorem. At last,
we investigate that all  product bases of $\mathbb{C}^{2}\bigotimes\mathbb{C}^{n}$ and obtain at
least $p(n)+1$ types in the last section. On the base of this paper, we will characterize
the structure
of product bases of $\mathbb{C}^{m}\bigotimes\mathbb{C}^{n}$ in sequel.

\section{The structure of product bases of $\mathbb{C}^2\bigotimes \mathbb{C}^n$}

In this section, we show that the first components and the second components of product bases of the bipartite system $\mathbb{C}^{2}\bigotimes\mathbb{C}^{n}$ can be grouped into $n$ orthonormal bases of $\mathbb{C}^2$ and 2 orthonormal bases of $\mathbb{C}^n$.

\begin{remark}
~Clearly, the product basis $B=\{|{a}_{i}\rangle\otimes|{b}_{i}\rangle~|~i=1,2,\ldots,d=mn\}$ of
 a complex vector space $\mathbb{C}^{m}\bigotimes\mathbb{C}^{n}$ satisfies the following
conditions: $\langle{a}_{i}|{a}_{j}\rangle=0$ or $\langle{b}_{i}|{b}_{j}\rangle=0$ for
any $i, j\in\{1, 2, \ldots, d\}$, $i\ne j$.
\end{remark}

\begin{lemma}
~Let $B=\{|{a}_{i}\rangle\otimes|{b}_{i}\rangle~|~i=1,2,\ldots,2n\}$
be a product basis of
 $\mathbb{C}^{2}\bigotimes\mathbb{C}^{n}$. Then for each  $|a_i\rangle$,
  $i\in\{1,2,\ldots,2n\}$, there exists  $|a_j\rangle$ such that $\langle a_{i}|a_{j}\rangle=0$.
\end{lemma}
\noindent{\textit{Proof.}}~~Suppose on the contrary. Then $\langle b_{i}|b_{j}\rangle=0$
for all $j$, $j\ne i$ and $j\in\{1,2,\ldots,2n\}$. Let $V$ and $W$ be the subspaces generated by
$|b_i\rangle$ and $\{|b_{j}\rangle~|~j\ne i, j\in\{1,2,\ldots,2n\}\}$.
Then $W=V^{\bot}$. Clearly,
 $|{a}_{i}\rangle\otimes|{b}_{i}\rangle\in \mathbb{C}^{2}\bigotimes V$ and
$|{a}_{j}\rangle\otimes|{b}_{j}\rangle\in \mathbb{C}^{2}\bigotimes W$ for all $j$, $j\ne i$ and $j\in\{1,2,\ldots,2n\}$. This shows that
$\{|{a}_{j}\rangle\otimes|{b}_{j}\rangle~|~j\ne i, j\in\{1,2,\ldots,2n\}\}$ is linearly dependent,
which is impossible.

Therefore, we obtain our result.\qed

By this lemma, the product basis $B=\{|{a}_{i}\rangle\otimes|{b}_{i}\rangle~|~i=1,2,\ldots,2n\}$ of
 $\mathbb{C}^{2}\bigotimes\mathbb{C}^{n}$ can be rearranged  in the following form:
\begin{center}
  $\{|{a}_{i}\rangle\otimes A(a_{i}), |a^{\bot}_{i}\rangle\otimes A(a^{\bot}_{i})~|~i=1, 2, \ldots, r\}$,
\end{center}
where $$A(a_{i})=\{|b_{j}\rangle~|~|a_{i}\rangle\otimes|b_{j}\rangle \in B,j=1,2,\ldots,2n\},$$
 $$A(a^{\bot}_{i})=\{|b_{j}\rangle~|~|a^{\bot}_{i}\rangle\otimes|b_{j}\rangle \in B,
j=1,2,\ldots,2n\}.$$

\begin{theorem}
~Given a product basis $B=\{|{a}_{i}\rangle\otimes A(a_{i}),
|a^{\bot}_{i}\rangle\otimes A(a^{\bot}_{i})~|~i=1, 2, \ldots, r\}$ as above. Then
the cardinalities of $A(a_{i})$ and $A(a^{\bot}_{i})$ are equal, i.e.,
 $|A(a_{i})|=|A(a^{\bot}_{i})|=m_{i}$, $i=1,2,\ldots,r$ and
$\sum\limits_{i=1}^{r}m_{i}=n$.
\end{theorem}

\noindent{\textit{Proof.}}~~
 According to the orthogonality, we find that the elements of $A(a_{i})$ are orthogonal
to the elements of $A(a_{j})$ and the elements of $A(a^{\bot}_{j})$ for any $i,j=1,2,\ldots,r$ and $i\ne j$. Besides, the elements of $A(a_{i})$ are mutually orthogonal and the elements of $A(a^{\bot}_{i})$ are also mutually orthogonal. Let $|A(a_{i})|=m_{i}$, $|A(a^{\bot}_{i})|=n_{i}$ for any $i=1,2,\ldots,r$. So, we obtain that
$\sum\limits_{i=1}^{r}m_{i}\leq n$
, $\sum\limits_{i=1}^{r}n_{i}\leq n$, $n_{j}+\sum\limits_{i\ne j}m_{i}\leq n$
and $m_j+\sum\limits_{i\ne j}n_{i}\leq n$ for every $j=1,2,\ldots,r$. On the other hand,
$\sum\limits_{i=1}^{r}m_{i}+\sum\limits_{i=1}^{r}n_{i}=2n$.
Thus, $\sum\limits_{i=1}^{r}m_{i}=\sum\limits_{i=1}^{r}n_{i}=n$. Let
$U_{i}$ and $V_{i}$ be the subspaces generated by $A(a_{i})$ and $A(a^{\bot}_{i})$ respectively.
Then
$\mathbb{C}^n=\bigoplus\limits_{i=1}^{r}U_{i}=\bigoplus\limits_{i=1}^{r}V_{i}$, where the symbol $\bigoplus\limits_{i=1}^{r}U_{i}$ means that $U_{i}$~$(i=1,2,\ldots,r)$ are mutually orthogonal. From the orthogonality,
we obtain $U_{i}\subseteq V_{i}$. Dually, $V_{i}\subseteq U_{i}$.
 So, $U_{i}=V_{i}$ for all $i=1, 2, \ldots, r$. Consequently, $m_{i}=n_{i}$ for all
$i=1, 2, \ldots, r$.\qed

Conversely, let $\mathbb{C}^n=\bigoplus\limits_{i=1}^{r}V_{i}$, $A_{i}$ and $B_{i}$
are the orthonormal bases of $V_{i}$, $|a_{i}\rangle$ and $|a^{\bot}_{i}\rangle$ are the states
in $\mathbb{C}^2$, where $|a^{\bot}_{i}\rangle$ is the unique state orthogonal to
 $|a_{i}\rangle$, $i=1, 2, \ldots, r$. Then, it is easy to check that the set
$\{|a_{i}\rangle\otimes A_{i}, |a^{\bot}_{i}\rangle\otimes B_{i}~|~i=1, 2, \ldots, r\}$
constitutes an orthonormal basis of $\mathbb{C}^2\bigotimes \mathbb{C}^n$. Combining  Lemma 2.2 with Theorem 2.3, we have the following result.

\begin{theorem}{\textbf{ (Main theorem)}}
~The set $B=\{|{a}_{i}\rangle\otimes|{b}_{i}\rangle~|~i=1,2,\ldots,d\}$ is an orthonormal
product basis of the space $\mathbb{C}^{2}\bigotimes\mathbb{C}^{n}$ if and only if the $d$ vectors
$|{a}_{i}\rangle~(i=1, 2, \ldots, d)$ and the $d$ vectors $|{b}_{i}\rangle~(i=1, 2, \ldots,d)$
can be grouped into $n$ orthonormal bases $B_{i}(2)$~$(i=1,2,\ldots,r)$ and $2$ orthonormal
bases $B_{i}(n)~(i=1,2)$, respectively, where $B_{i}(2)=\{|{a}_{i}\rangle,|{a}^{\bot}_{i}\rangle\}$,
$|B_{i}(2)|=m_{i}~(i=1,2,\ldots,r)$, $B_{1}(n)=\bigcup\limits_{i=1}^{r} A(a_{i})$,
$B_{2}(n)=\bigcup\limits_{i=1}^{r} A(a^{\bot}_{i})$, $A(a_{i})$ and
$A(a^{\bot}_{i})$ are the orthonomal bases of the subspace $V_{i}$ and $\mathbb{C}^n=\bigoplus\limits_{i=1}^{r}V_{i}$.
\end{theorem}

\begin{definition}
~Let $\mathbb{C}^2\bigotimes \mathbb{C}^n$ be a bipartite system.
If $\mathbb{C}^{n}=\bigoplus\limits_{i=1}^{r}V_{i}$ and
$dimV_{i}=m_{i}$ for $i=1, 2, \ldots, r$, then $(m_{1}, m_{2}, \ldots, m_{r})$
is said to be a right type of $\mathbb{C}^2\bigotimes \mathbb{C}^n$.
\end{definition}

\section{The product bases of $\mathbb{C}^{2}\bigotimes\mathbb{C}^{2}$ and
$\mathbb{C}^{2}\bigotimes\mathbb{C}^{3}$}

McNulty and Weigert \cite{6} discussed all the product bases of a complex vector
space in dimension 4 and 6. However, it is difficult for us to go further in
 high dimension spaces by their's method. Moreover, they also proposed the definition of local equivalent transformations{\textbf{(LETs)}. LETs are defined by the requirement that they preserve the product structure of all states. In this section,  we
review the product bases of $\mathbb{C}^{2}\bigotimes\mathbb{C}^{2}$ and
$\mathbb{C}^{2}\bigotimes\mathbb{C}^{3}$ from the perspective of Main theorem.
\subsection{The product bases of $\mathbb{C}^{2}\bigotimes\mathbb{C}^{2}$}

\begin{lemma}{\textbf{(Lemma 1 \cite{6})}}
~Any orthonormal product basis of the space $\mathbb{C}^{2}\bigotimes\mathbb{C}^{2}$ is
equivalent to a member of one of the families
\vskip3mm
\hskip15mm
$I_{0}=\{|j_{z},k_{z}\rangle\}$,
\vskip3mm
\hskip15mm
$I_{1}=\{|0_{z},k_{z}\rangle, |1_{z},\widehat{u}k_{z}\rangle\}$,
\vskip3mm
\hskip15mm
$I_{2}=\{|j_{z},0_{z}\rangle, |\widehat{v}j_{z},1_{z}\rangle\}$,

\hskip-7mm
where the operators $\widehat{u}$, $\widehat{v}\in SU(2)$ act on the space $C^2$ such
that the states $|0_{z}\rangle$ and $\widehat{u}|0_{z}\rangle$, as well as the states
$|0_{z}\rangle$ and $\widehat{v}|0_{z}\rangle$, are skew.
\end{lemma}

McNulty and Weigert's work are  a bit not concise. Next, we apply the Main theorem
 to describe all the product bases in dimension 4. Any orthonormal product basis of the space
 $\mathbb{C}^{2}\bigotimes\mathbb{C}^{2}$ must be of the form
$\{|a_{i}\rangle\otimes|b_{i}\rangle~|~i=1, 2, 3, 4\}$. From the Theorem 2.4,
we know that $\{|a_{i}\rangle~|~i=1, 2, 3, 4\}$ can be grouped into 2 orthonormal bases
of $\mathbb{C}^2$. Therefore, the product basis must be one of the following cases:

\textbf{Case 1.}$\{|a_{1}\rangle\otimes|b_{1}\rangle,|a^{\bot}_{1}\rangle\otimes|c_{1}\rangle,
|{a}_{2}\rangle\otimes|{b}_{2}\rangle,|{a}^{\bot}_{2}\rangle\otimes|{c}_{2}\rangle\}$,
 where $|a_{1}\rangle\neq |a_{2}\rangle$ and $\langle a_{1}|a_{2}\rangle\neq 0$.
In this case, from the orthogonality, we obtain that
 $|l_{1}\rangle=|l^{\bot}_{2}\rangle$~$(l=b, c)$ and $|b_{j}\rangle=|c_{j}\rangle$~$(j=1, 2)$.
 Thus, the product basis is of the form
\begin{center}
$B_{0}=\{|a_{1}\rangle\otimes|b_{1}\rangle,|a^{\bot}_{1}\rangle\otimes|b_{1}\rangle,
|a_{2}\rangle\otimes|b^{\bot}_{1}\rangle,|a^{\bot}_{2}\rangle\otimes|b^{\bot}_{1}\rangle\}$.
\end{center}

\textbf{Case 2.}$\{|{a}\rangle\otimes\{|{b}_{1}\rangle,|{b}^{\bot}_{1}\rangle\},
|{a}^{\bot}\rangle\otimes\{|{b}_{2}\rangle,|{b}^{\bot}_{2}\rangle\}\}$. If
$\{|{b}_{1}\rangle,|{b}^{\bot}_{1}\rangle\}$ is different from
$\{|{b}_{2}\rangle,|{b}^{\bot}_{2}\rangle\}$, we obtain that the product basis is of the form
\begin{center}
$B_{1}=\{|{a}\rangle\otimes|{b}_{1}\rangle,|{a}\rangle\otimes|{b}^{\bot}_{1}\rangle,
|{a}^{\bot}\rangle\otimes|{b}_{2}\rangle,|{a}^{\bot}\rangle\otimes|{b}^{\bot}_{2}\rangle\}.$
\end{center}
Otherwise, we have the product basis is a direct product basis, which is
\begin{center}
$B_{2}=\{|{a}\rangle\otimes|{b}\rangle,|{a}\rangle\otimes|{b}^{\bot}\rangle,
|{a}^{\bot}\rangle\otimes|{b}\rangle,|{a}^{\bot}\rangle\otimes|{b}^{\bot}\rangle\}.$
\end{center}

The right type is (1,1) in case 1 and the right type is (2,0) in case 2. Because of
the same right type of $B_{1}$ and $B_{2}$, we can consider the left type similarly.
The left type of $B_{1}$ is (1,1), but the left type of $B_{2}$ is (2,0). We also know
 that the left type of $B_{0}$ is (2,0). The types of the construction given above are the same as McNulty and Weigert's work.\qed

From  Corollary 1 of \cite{3},
we know that, up to local equivalence transformations, there exists a unique triple of MUPBs in
dimension 4 as follows,
\vskip3mm
\hskip15mm
$B'_{0}=\{|{j}^{1}_{z}\rangle\otimes|{j}^{2}_{z}\rangle\}$,
\vskip3mm
\hskip15mm
$B'_{1}=\{|{j}^{1}_{x}\rangle\otimes|{j}^{2}_{x}\rangle\}$,
\vskip3mm
\hskip15mm
$B'_{2}=\{|{j}^{1}_{y}\rangle\otimes|{j}^{2}_{y}\rangle\}$,

\hskip-7mm
here $\{|j^{r}_{b}\rangle~|~j=0,1\}, b=z,x,y$,
are, for each $r=1,2$, the eigenstates of the three Pauli operators in $\mathbb{C}^2$.
The three MUPBs are corresponding to the product basis $B_{2}$.

\subsection{The product bases of $\mathbb{C}^{2}\bigotimes\mathbb{C}^{3}$}

\begin{lemma}{\textbf{ (Lemma 2 \cite{6})}}
~Any orthonormal product basis of the space $\mathbb{C}^{2}\bigotimes\mathbb{C}^{3}$ is
equivalent to a member of one of the families
\vskip3mm
\hskip15mm
$I_{0}=\{|j_{z},J_{z}\rangle\}$,
\vskip3mm
\hskip15mm
$I_{1}=\{|0_{z},J_{z}\rangle, |1_{z},\widehat{U}J_{z}\rangle\}$,
\vskip3mm
\hskip15mm
$I_{2}=\{|j_{z},0_{z}\rangle, |\widehat{u}0_{z},1_{z}\rangle, |\widehat{u}0_{z},2_{z}\rangle,
 |\widehat{u}1_{z},\widehat{V}1_{z}\rangle, |\widehat{u}1_{z},\widehat{V}2_{z}\rangle\}$,
\vskip3mm
\hskip15mm
$I_{3}=\{|j_{z},0_{z}\rangle, |\widehat{v}j_{z},1_{z}\rangle, |\widehat{w}j_{z},2_{z}\rangle\}$,

\hskip-7mm
with $j=0, 1$ and $J=0, 1, 2$; the operators $\widehat{u}, \widehat{v}, \widehat{w}\in SU(2)$
and $\widehat{U}, \widehat{V}\in SU(3)$ act on $C^2$ and $C^3$, respectively, with
$\widehat{V}$ leaving the state $|0_{z}\rangle$ invariant; the parameters of the operators
$\widehat{u}, \ldots, \widehat{V}$ are chosen in such a way that no product basis occurs more
than once.
\end{lemma}

We  next describe all the
product bases in dimension 6 by the Main theorem. The procedure and results are a slight different
from  the research by McNulty and Weigert. Consider an orthonormal product
 basis of the complex vector space $\mathbb{C}^{2}\bigotimes\mathbb{C}^{3}$:

\textbf{Case 1. The right type is (1,1,1)}:
The form of the product basis must be
$\{|{a}_{1}\rangle\otimes|{b}_{1}\rangle,|{a}^{\bot}_{1}\rangle\otimes|{c}_{1}\rangle,
|{a}_{2}\rangle\otimes|{b}_{2}\rangle,|{a}^{\bot}_{2}\rangle\otimes|{c}_{2}\rangle,
|{a}_{3}\rangle\otimes|{b}_{3}\rangle,|{a}^{\bot}_{3}\rangle
\otimes|{c}_{3}\rangle\}$, where $|a_{i}\rangle\neq |a_{j}\rangle$ and
$\langle a_{i}|a_{j}\rangle\neq 0$ for any $i, j\in\{1, 2, 3\}$ and $i\neq j$.
The orthogonality condition implies that both $\{|{b}_{1}\rangle,|{b}_{2}\rangle,|{b}_{3}\rangle\}$
and $\{|{c}_{1}\rangle,|{c}_{2}\rangle,|{c}_{3}\rangle\}$ are orthonormal bases of $\mathbb{C}^{3}$.
Since for each $i$, $i=1,2,3$ the subspaces generated by  $|{b}_{i}\rangle$ and $|{c}_{i}\rangle$
 are the same one, it follows that $|{b}_{i}\rangle=|{c}_{i}\rangle$.

The form of this kind basis is as follows:

\begin{center}
 $B_{0}=\{|{a}_{1}\rangle\otimes|{b}\rangle,|{a}^{\bot}_{1}\rangle\otimes|{b}\rangle,
|{a}_{2}\rangle\otimes|{b}^{\bot}\rangle,|{a}^{\bot}_{2}\rangle\otimes|{b}^{\bot}\rangle,
|{a}_{3}\rangle\otimes|{b}^{\bot\bot}\rangle,
|{a}^{\bot}_{3}\rangle\otimes|{b}^{\bot\bot}\rangle\}$.
\end{center}

\textbf{Case 2. The right type is (2,1)}:
The form of the product basis must be:
$\{|{a}_{1}\rangle\otimes\{|{b}_{1}\rangle,|{b}^{\bot}_{1}\rangle\},
|{a}^{\bot}_{1}\rangle\otimes\{|{b}_{2}\rangle,|{b}^{\bot}_{2}\rangle\},
|{a}_{2}\rangle\otimes|{b}_{3}\rangle,|{a}^{\bot}_{2}\rangle\otimes|{b}_{4}\rangle\}$,
 where $\{|a_{1}\rangle,|{a}^{\bot}_{1}\rangle\}$ and
$\{|a_{2}\rangle,|{a}^{\bot}_{2}\rangle\}$ are different orthonormal bases of a subspace
$V_1$ in dimension 2, $|{b}_{3}\rangle$ and $|{b}_{4}\rangle$ generate the same subspace $V_2$ in dimension 1, $V_1$ and $V_2$ are orthogonal.

Thus, the product bases is of the following form (consults \cite{6}):

\begin{center}
$B_{1}=\{|{a}_{1}\rangle\otimes|b\rangle,|{a}_{1}\rangle\otimes|b^{\bot}\rangle,
|{a}^{\bot}_{1}\rangle\otimes
\widehat{V}|b\rangle,|{a}^{\bot}_{1}\rangle\otimes \widehat{V}|{b}^{\bot}\rangle,
|{a}_{2}\rangle\otimes|{b}^{\bot\bot}\rangle,|{a}^{\bot}_{2}\rangle
\otimes|{b}^{\bot\bot}\rangle\}$,
\end{center}
where $\widehat{V}$ is a unitary operator of $V_1$ defined by $\widehat{V}|b\rangle=
\alpha|b\rangle+\beta|b^{\bot}\rangle$,~$\widehat{V}|b^{\bot}\rangle=
\overline{\beta}|b\rangle-\overline{\alpha}|b^{\bot}\rangle$,~$|\alpha|^2+|\beta|^2=1$.

\textbf{Case 3. The right type is (3,0)}: In this case, the product basis is\\
$$\{|a\rangle\otimes\{|{b}_{1}\rangle,
|{b}^{\bot}_{1}\rangle,|{b}^{\bot\bot}_{1}\rangle\},|{a}^{\bot}\rangle\otimes\{|b_{2}\rangle,
|{b}^{\bot}_{2}\rangle,
|{b}^{\bot\bot}_{2}\rangle\}\}.$$
\hskip-1mm
If $\{|{b}_{1}\rangle,|{b}^{\bot}_{1}\rangle,
|{b}^{\bot\bot}_{1}\rangle\} \neq\{|b_{2}\rangle,|{b}^{\bot}_{2}\rangle,
|{b}^{\bot\bot}_{2}\rangle\}$, then the product basis is of the form
\begin{center}
$B_{2}=\{|a\rangle\otimes\{|{b}_{1}\rangle,|{b}^{\bot}_{1}\rangle,|{b}^{\bot\bot}_{1}\rangle\},
|{a}^{\bot}\rangle\otimes
\{|b_{2}\rangle,|{b}^{\bot}_{2}\rangle,|{b}^{\bot\bot}_{2}\rangle\}\}$.
\end{center}
If it is not so, then the product basis is a direct product basis
\begin{center}
$B_{3}=\{|a\rangle\otimes\{|{b}\rangle,|{b}^{\bot}\rangle,|{b}^{\bot\bot}\rangle\},
|{a}^{\bot}\rangle\otimes\{|b\rangle,|{b}^{\bot}\rangle,
|{b}^{\bot\bot}\rangle\}\}$.
\end{center}

If we look into the left type in case 2, then we  find that there exist some difficulties.
 If $\widehat{V}|b\rangle=|b\rangle$, $\widehat{V}|b^{\bot}\rangle=|b^{\bot}\rangle$ or
$\widehat{V}|b\rangle=|b^{\bot}\rangle$, $\widehat{V}|b^{\bot}\rangle=|b\rangle$, then the
left type of $B_{1}$ is (2,0). If $\widehat{V}|b\rangle$ is not so, then the left type of $\widehat{V}|b\rangle$ is (1,1).
 However, the left type of $|{b}^{\bot\bot}\rangle$ is (2,0).
In case 3, the left type corresponding $B_{2}$ is not well-defined. Due to this fact,
 we only consider the right type of $\mathbb{C}^{2}\bigotimes\mathbb{C}^{3}$.\qed

It is still unknown if there exist more than three MUBs of a complex vector space with
dimension 6. A possible choice\cite{7} of three MUBs is to take the products
$B'_{0}={I}_{6}={I}_{2}\otimes{I}_{3}$, $B'_{1}={B}_{11}\otimes{B}_{12}$,
$B'_{2}={B}_{21}\otimes{B}_{22}$, where ${I}_{2}$ and ${I}_{3}$ denote identity operators and
\begin{equation*}
B_{11}=\frac{1}{\sqrt{2}}
\begin{pmatrix} 1 & 1 \\ 1 & -1 \end{pmatrix}, B_{12}=\frac{1}{\sqrt{3}}
\begin{pmatrix} 1 & 1 & 1 \\ 1 & \omega & {\omega}^2 \\ 1 & {\omega}^2 & \omega \end{pmatrix}
\end{equation*}

\begin{equation*}
B_{21}=\frac{1}{\sqrt{2}}
\begin{pmatrix} 1 & 1 \\ i & -i \end{pmatrix}, B_{22}=\frac{1}{\sqrt{3}}
\begin{pmatrix} 1 & 1 & 1 \\ \omega & {\omega}^2 & 1 \\ \omega & 1 & {\omega}^2 \end{pmatrix}
\end{equation*}
with $\omega=e^{\frac{2\pi{i}}{3}}$. The three MUBs are corresponding to the product basis $B_{3}$.

\section{The product bases of $\mathbb{C}^{2}\bigotimes\mathbb{C}^{n}$}

The analysis in Section 2 indicates that the right type of $\mathbb{C}^{2}\bigotimes\mathbb{C}^{n}$ is
 correspongding to a partition of the positive integer $n$. A partition of a positive integer
 $n$ is a representation of writing $n$ as a sum of several positive integers. That is,
 if $n_{1}, n_{2}, \ldots, n_{k}$ are positive integers and
$n_{1}\geq n_{2}\geq \cdots \geq n_{k}$, then the representation
$n=n_{1}+n_{2}+\cdots+n_{k}$ is called a partition of $n$ with $k$ parts.
If there have no other restrictions on $n_{i}$ and $k$, then we call it an unrestricted partition
or partition simply.
Let $p(n)$ denote the number of unrestricted partitions of $n$. Then we have the following result.

\begin{theorem}
~A product basis of the complex vector space $\mathbb{C}^{2}\bigotimes\mathbb{C}^{n}$ is corresponding
a partition of $n$ and vice versa.
\end{theorem}
\noindent{\textit{Proof.}}~~By the Main theorem, if $\mathbb{C}^n=\bigoplus\limits_{i=1}^{r}V_{i}$, $A(a_{i})$ and
$A(a^{\bot}_{i})$ are the orthonomal bases of $V_{i}$ for each $i$, then
 $\{|a_{i}\rangle\otimes A(a_{i}),|a^{\bot}_{i}\rangle\otimes A(a^{\bot}_{i})~|~i=
1, 2, \ldots, r\}$ is the product basis of $\mathbb{C}^{2}\bigotimes\mathbb{C}^{n}$,
where $|a_{i}^{\bot}\rangle$ is the unique state orthogonal to $|a_{i}\rangle$ for any $i=1,2,\ldots,r$. We  obtain that an orthonormal product basis of the complex
vector space $\mathbb{C}^{2}\bigotimes\mathbb{C}^{n}$ as follows:
\begin{center}
$B=\{|a_{i}\rangle\otimes A(a_{i}),|a^{\bot}_{i}\rangle\otimes A(a^{\bot}_{i})~|~i=
1, 2, \ldots, r\}$
\end{center}
where $|A(a_{i})|=|A(a^{\bot}_{i})|=n_{i}$ and $\sum\limits_{i=1}^{r}n_{i}=n$,~$1\le r\le n$.
In other words, given a partition of $n$, there exists a corresponding product basis. \qed

Clearly, $n_{1}=n$ is a partition of $n$. If we chose two orthogonal bases $B_1$ and $B_2$
of the complex vector space
$\mathbb{C}^{n}$, then $B_1=B_2$ or $B_1\ne B_2$. The first case yields
a direct product basis of $\mathbb{C}^2\bigotimes \mathbb{C}^n$. So, we have

\begin{theorem}
~There are at least two  product bases of the complex vector space
$\mathbb{C}^{2}\bigotimes\mathbb{C}^{n}$ whose right type is $(n,0)$.
\end{theorem}

Furthermore, we have

\begin{theorem}~Let $(m_{1}, m_{2}, \ldots, m_{r})$
be a right type of $\mathbb{C}^2\bigotimes \mathbb{C}^n$ and $\mathbb{C}^n=\bigoplus\limits_{i=1}^{r}V_{i}$. If $m_i\geq 1$, then
there exist two product bases $A^{k_i}_i$ $(k_i=1,2)$ of the complex vector space
$\mathbb{C}^{2}\bigotimes V_i$ such that $\{A^{k_i}_i~|~i=1,2,\ldots,r\}$ forms a
product basis of $\mathbb C^2\bigotimes\mathbb C^n$.
\end{theorem}

The analysis indicates that there is no product basis of $\mathbb C^2\bigotimes\mathbb C^n$
whose  the left type is (1,1)
and the right type is $\underbrace{(1,1,\ldots,1)}_{n}$. That is, if all of $n$ bases are
different, then
\begin{center}
$B=\{|a_{i}\rangle\otimes|{A^{(i)}}\rangle,|a^{\bot}_{i}\rangle\otimes|{B^{(i)}}\rangle~|~i=
1,2,\ldots,n\}$,
\end{center}
where $|a_{i}^{\bot}\rangle$ is the unique state orthogonal to $|a_{i}\rangle$ for $i=
1,2,\ldots,n$, both $\{|{A^{(i)}}\rangle~|~i=1,2,\ldots,n\}$ and $\{|B^{(i)}\rangle~|~i=
1,2,\ldots,n\}$ are the orthonormal bases of ${C}^{n}$.
From the orthogonal conditions, we obtain that $\{|{A^{(i)}}\rangle~|~i=1,2,\ldots,n\}
=\{|B^{(i)}\rangle~|~i=1,2,\ldots,n\}$. Therefore, if we consider the left type of $B$,
then the left type of $B$ must be (2,0). In other words, there must exist product
bases of the form of the left type is (2,0) and the right type is
$\underbrace{(1,1,\ldots,1)}_{n}$.

How many different product bases are there in $\mathbb C^2\bigotimes\mathbb C^n$?
A rough estimate answer is, there are at least $p(n)+1$ types.

From  Corollary 4 of \cite{3}, we know that any triple of MUPBs of the complex vector space
 $\mathbb{C}^{2}\bigotimes\mathbb{C}^{n}$ must have the following form
\vskip3mm
\hskip15mm
$B'_{0}=\{|{j}_{z}\rangle\otimes|G({j}_{z})\rangle\}$,
\vskip3mm
\hskip15mm
$B'_{1}=\{|{j}_{x}\rangle\otimes|G({j}_{x})\rangle\}$,
\vskip3mm
\hskip15mm
$B'_{2}=\{|{j}_{y}\rangle\otimes|G({j}_{y})\rangle\}$,

\hskip-7mm
up to local equivalence transformations, here $\{|j_{b}\rangle~|~j=0,1\}, b=z,x,y$, are the eigenstates of the three Pauli operators of $\mathbb{C}^2$, and $G({j}_{b})$ are
bases of $\mathbb{C}^{n}$ for each $j_{b}$, such that the three set $\{G({j}_{b})~|~j=0,1\}$
 are mutually unbiased.
The three MUPBs are corresponding to the direct product basis.


\begin{thebibliography}{99}
\bibitem{1} A. Einstein, B. Podolsky and N. Rosen, Phys. Rev. \textbf{47}, 777 (1935)

\bibitem{2} N. Alon and L. Lov\'{a}sz, Unextendible product bases. Journal of Combinatorial
Theory, Series A, \textbf{95}(1), 169-179 (2001)

\bibitem{3} D. McNulty, B. Pammer and S. Weigert, Mutually Unbiased Product Bases for
Multiple Qudits. Journal of Mathematical Physics, \textbf{57}, 032202 (2016)

\bibitem{4} W. K. Wootters and B. D. Fields, Optimal state-determination by mutually
unbiased measurements, Ann. Physics, \textbf{191}, 363-381, (1989)

\bibitem{5} A. Klappenecker and M. R\"{o}tteler, Finite Fields and Applications,
Lecture Notes in Computer Science Vol. 2948 (Springer, Berlin, 2004), pp. 137-144.

\bibitem{6} D. McNulty and S. Weigert, All mutually unbiased product bases in
dimension 6. Journal of Physics A: Mathematical and Theoretical \textbf{45}(13),
 135307 (2012)

\bibitem{7} M. Wie\'{s}niak, T. Paterek and A. Zeilinger, Entanglement in mutually
unbiased bases, New Journal of Physics, \textbf{13}(5), 053047 (2011)

\bibitem{8} D. P. DiVincenzo and B. M. Terhal, Product Bases in Quantum Information Theory [R]. \textbf(2000)
\end{thebibliography}
\end{document}